# Evidence of Overcharging in the Complexation between Oppositely Charged Polymers and Surfactants

**Jean-François Berret**[@]
Complex Fluids Laboratory, CNRS - Cranbury Research Center Rhodia Inc.,
259 Prospect Plains Road CN 7500, Cranbury NJ 08512 USA

**Abstract :** We report on the complexation between charged-neutral block copolymers and oppositely charged surfactants studied by small-angle neutron scattering. Two block copolymers/surfactant systems are investigated, poly(acrylic acid)-*b*-poly(acrylamide) with dodecyltrimethylammonium bromide and poly(trimethylammonium ethylacrylate methylsulfate)-*b*-poly(acrylamide) with sodium dodecyl sulfate. The two systems are similar in terms of structure and molecular weight but have different electrostatic charges. The neutron scattering data have been interpreted in terms of a model that assumes the formation of mixed polymer-surfactant aggregates, also called colloidal complexes. These complexes exhibit a core-shell microstructure, where the core is a dense coacervate microphase of micelles surrounded by neutral blocks. Here, we are taking advantage of the fact that the complexation results in finite-size aggregates to shed some light on the complexation mechanisms. In order to analyze quantitatively the neutron data, we develop two different approaches to derive the number of surfactant micelles per polymer in the mixed aggregates and the distributions of aggregation numbers. With these results, we show that the formation of the colloidal complex is in agreement with overcharging predictions. In both systems, the amount of polyelectrolytes needed to build the core-shell colloids always exceeds the number that would be necessary to compensate the charge of the micelles. For the two polymer-surfactant systems investigated, the overcharging ratios are $0.66 \pm 0.06$ and $0.38 \pm 0.02$.

# I - Introduction

The complexation between ion-containing polymers and oppositely charged macroions is currently attracting much attention in the field of soft condensed matter[1,2]. Associations based on the electrostatic Coulomb interactions are present in many applications and for highly charged systems the elementary mechanisms such as adsorption and self-assembly are only partially understood. The complexation between oppositely charged species is found e.g. in the treatment of waste water, the formulation of personal care products, the purification of proteins or in gene and drug delivery. In biophysics, comprehensive studies have been dedicated to the cooperative condensation of DNA with multivalent counterions[3-5] or with cationic liposomes[1,6-9]. In material science, electrostatic layer-by-layer deposition yielding polyelectrolyte multilayers have been achieved for encapsulation purposes and colloidal stabilization[10-12].

When an ion-containing polymer solution is mixed to a dispersion of oppositely charged colloids, a phase separation usually follows[13-23]. At the mixing, the solution becomes turbid, and after centrifugation it displays two separated phases. The bottom phase appears as a precipitate and its chemical analysis reveals that it contains most of the polymers and colloids. The supernatant is fluid and transparent, and contains the solvent and the counterions released during the separation. The mixing conditions of such experiments are that *i)* the ratio Z between the electrostatic charges borne by the colloids and the polymers is of the order of unity, and *ii)* the overall concentration is typical of the concentrations used in applications, with the additional constraint that the salt content remains low. In recent years, the mechanisms proposed for such a phase separation have been the subject to intense debates. For the DNA strands complexed with cationic liposomes, Wagner *et al.*[8] have found direct evidences that the complexation is dominated by the entropic contribution resulting from counterion release. Here, experiments and theory agree well with each other to show a maximal and complete release of the monovalent counterions at the isoelectric point, $Z = 1$[1,6-8]. Other approaches are also based on the mechanism of counterion release, but additional hypothesis have been proposed. This is the case of the models predicting overcharging[24-29]. When a polyelectrolyte adsorbs on an oppositely charged surface, that of a colloid for instance, the positive and negative charges do not compensate. The amount of polymers involved in the adsorption process is such that the initial charge of the colloid can be reversed. Monte Carlo simulations on a polyelectrolyte-colloid pair have shown that overcharging can be observed when the polymer is large enough[26,27,30]. Extending the previous models, Grosberg and coworkers suggested recently that multivalent ions adsorbed on a spherical macroion can be strongly correlated and build the analogue of a two dimensional liquid[26,31,32]. For polymers, the chains wind around the macroions and form an almost equidistant "solenoid"[26]. This correlated structure can induce attractions between macroions and finally cause a phase separation[33].

In the present paper, we investigate the complexation between charged-neutral block copolymers and oppositely charged surfactants by means of small-angle neutron scattering (SANS). Two copolymer-surfactant systems are put under scrutiny, the anionic-neutral copolymer poly(acrylic acid)-*b*-poly(acrylamide) with dodecyltrimethylammonium bromide[34,35] and the cationic-neutral copolymer poly(trimethylammonium ethylacrylate methylsulfate)-*b*-poly(acrylamide) with sodium dodecyl sulfate[36,37]. The two systems are very similar in terms of molecular weights and structures but they differ by their electrostatic charge. For charged-neutral diblocks and surfactants with opposite charges, we have established the thermodynamical phase diagrams and found broad regions where hierarchical aggregates spontaneously form via electrostatic self-assembly[34-40]. These aggregates are termed colloidal complexes because they form through complexation and because of their colloidal nature[21,41]. The surfactant-polymer assemblies exhibit a core-shell structure illustrated in Fig. 1. The core is a dense and disordered microphase made of surfactant micelles connected by the polyelectrolyte blocks. The corona is a diffuse shell of the neutral chains and it insures steric stability. Here, we are taking advantage of the fact that the complexation is controlled, resulting in finite-size aggregates to shed some light on the complexation mechanisms. This is achieved through an accurate analysis of the SANS data in terms of aggregation numbers and aggregation number distributions. Doing so, we are able to show that the formation of the colloidal complex is in agreement with overcharging predictions. For both systems, the amount of polyelectrolytes needed to build the core-shell structures always exceed the number that would be necessary to compensate the bare charges of the micelles.

## II - Theory

*Scattering by a distribution of homogeneous and spherical particles*

The neutron scattering cross-section of a dispersion of spherical colloids at volume fraction $\phi = n(\phi)V$ reads[42,43]:

$$\frac{d\sigma_M}{d\Omega}(q,\phi) = n(\phi)\Delta\rho^2 V^2 S_M(q,\phi) F(q,R) \tag{1}$$

where $n(\phi)$ is the number density of spheres of volume V and radius R, $\Delta\rho$ is the scattering length density difference with respect to the solvent. $F(q,R)$ and $S_M(q,\phi)$ are the form and structure factors of the spheres. For homogeneous spheres, $F(q) = [3(\sin X - X\cos X)/X^3]^2$ where $X = qR$. To keep the notations consistent throughout the paper, we index the cross-section and the structure factor in Eq. 1 with the letter M, for micelle. At sufficiently low $\phi$, the interparticle interactions are negligible and $S_M(q,\phi) = 1$. In this regime and for particles distributed according to a Gaussian distribution function, the scattering cross-section becomes:

$$\frac{d\sigma_M^{(d)}}{d\Omega}(q,\phi) = n(\phi)\Delta\rho^2 \int_0^\infty V^2 G(R,\overline{R},\sigma) F(q,R) dR \tag{2}$$

where the index d (here and below) reminds that the function is integrated over the distribution $G(R,\overline{R},\sigma_R)$. In Eq. 2, $\overline{R}$ and $\sigma_R$ are the average and standard deviation of the Gaussian function,

$$G(R,\overline{R},\sigma_R) = \frac{1}{\sigma_R\sqrt{2\pi}} \exp\left[-\left(\frac{(R-\overline{R})^2}{2\sigma_R^2}\right)\right]. \tag{3}$$

The polydispersity s is defined as $s = \sigma_R/\overline{R}$. As $q \to 0$ in Eq. 2, the scattering intensity reaches asymptotically a constant value, $n(\phi)\Delta\rho^2\overline{V^2}$, which can be rewritten as:

$$\frac{d\sigma_M^{(d)}}{d\Omega}(q \to 0,\phi) = \phi\Delta\rho^2 \frac{\overline{V^2}}{\overline{V}}. \tag{4}$$

For polydisperse spheres, the volume fraction is given by $\phi = n(\phi)\overline{V}$, where the upper bar indicates the summation over the distribution $G(R,\overline{R},\sigma_R)$.

*Scattering by clusters of spheres*

We now assume that the previous spherical particles have been assembled into clusters or aggregates, as illustrated in Fig. 2. We first consider the case where the small spheres are monodisperse. The clusters are characterized by an aggregation number $N \gg 1$ and a radius $R_C$ (Fig. 2). These two quantities are related by the expression :

$$N = \phi_C \left(\frac{R_C}{R}\right)^3 \tag{5}$$

where $\phi_C$ is the volume fraction of particles within a N-aggregate (or N-cluster). In the present work, the clusters are dense and $\phi_C$ is of the order of 0.4 - 0.5. We also assume that the structure of the cluster is disordered and that there is no positional long range order between the center-of-masses of the particles[37]. For a N-cluster, we denote $\mathbf{r_i}$ (with i = 1 to N) the position of each small spheres with respect to an arbitrarily chosen origin. The scattering intensity arising from such an aggregate was determined for the first time by Debye[44-46]:

$$\frac{d\sigma_C}{d\Omega}(q,N) = \Delta\rho^2 V^2 F(q,R) \left[ N + 2 \sum_{i=1}^{N-1} \sum_{j=i+1}^{N} \frac{\sin qr_{ij}}{qr_{ij}} \right] \tag{6}$$

where $r_{ij} = |\mathbf{r_i} - \mathbf{r_j}|$ is the distance between the pairs of spheres. In Eq. 6, also called the Debye Formula, the scattering depends on the choice of the positions $r_i$. Each aggregate with the same radius and aggregation number but with another distribution of spheres will have a slightly different scattering spectrum.

Our goal here is to use the Debye Formula to predict the scattering intensity of a dispersion of aggregates having the same radius $R_C$ and aggregation number N, but different configurations in the $r_i$'s. For a dispersion containing $n(\phi)$ small spheres per unit volume, and arranged into $n(\phi)/N$ clusters, Eq. 6 can be rewritten as[36,47] :

$$\frac{d\sigma_C}{d\Omega}(q,N,\phi) = n(\phi)\Delta\rho^2 V^2 F(q,R) \langle S_C(q,N) \rangle_{MonteCarlo}, \tag{7}$$

where

$$S_C(q,N) = 1 + \frac{2}{N} \sum_{i=1}^{N-1} \sum_{j=i+1}^{N} \frac{\sin qr_{ij}}{qr_{ij}}. \tag{8}$$

The brackets with the index "Monte Carlo" in the right hand side of Eq. 7 mean that the averaging over all positional configurations has been performed by Monte Carlo simulations. The Monte Carlo algorithm consists in letting the particles move by Brownian motions within the bounds of a spherical cluster and to calculate the structure factor $S_C(q,N)$ every 10N Monte Carlo steps[36,47]. Each Monte Carlo step is an infinitesimal displacement of a unique small sphere. This function is averaged over time, until this average reaches the stationary function $\langle S_C(q,N)\rangle_{MonteCarlo}$. Then, instead of averaging over clusters with different positional arrangements, we average here over the time during which the Brownian spheres explore a large number of configurations. In the ergodic approximation, these two averaging procedures are equivalent. It is interesting to note that the low wave-vector limit of $S_C(q,N)$ is N, *i.e.* the intensity of a dispersion of N-aggregates is N times that of a dispersion of unassociated small spheres.

*Scattering by polydisperse clusters*

For polydisperse clusters, we assume that $R_C$ also obeys a Gaussian distribution function[34]. Through Eq. 5, this distribution can be transformed into a distribution of aggregation numbers, noted $P(N,\overline{N},\sigma_N)$ where $\overline{N}$ is the average aggregation number and $\sigma_N = \sqrt{\overline{N^2}-\overline{N}^2}$ standard deviation. The average interparticle structure factor of Eq. 7 is now replaced by the expression[36]:

$$\langle S_C^{(d)}(q,N)\rangle_{MonteCarlo} = \frac{1}{\overline{N}}\int_0^\infty N\, P(N,\overline{N},\sigma_N)\langle S_C(q,N)\rangle_{MonteCarlo}\,dN \qquad (9)$$

In this expression as in Eq. 7, we recall that $\langle S_C(q,N)\rangle_{MonteCarlo}$ is the structure factor calculated for a N-aggregate using the Monte Carlo scheme described previously. The structure factor at zero scattering angle is now equal to $\overline{N^2}/\overline{N}$, instead of N for monodisperse spheres. For polydisperse clusters made from polydisperse spheres, we assume within the present model that the integration variables R and N can be separated. This approximation is justified experimentally and should hold as long as the two polydispersities are not too broad, say $s_R$, $s_{RC}$ < 0.2. In this case, quantitative fitting are achieved using the expression :

$$\frac{d\sigma_C^{(d)}}{d\Omega}(q,N,\phi) = \frac{d\sigma_M^{(d)}}{d\Omega}(q,\phi)\langle S_C^{(d)}(q,N)\rangle_{MonteCarlo} \qquad (10)$$

Note that in the Monte Carlo simulations, the small spheres selected to simulate a N-cluster are themselves polydisperse and in agreement with Eq. 3.

# III – Experimental and Results

## III.1 – Material and Characterization

In the present survey, we report on two block copolymers/surfactant systems[34-37]. The first system is the anionic-neutral copolymer poly(acrylic acid)-*b*-poly(acrylamide) which is studied in aqueous solutions with a positively charged surfactant, dodecyltrimethylammonium bromide (DTAB). The second system is reverse in terms of electrostatic charges. It is made of a cationic-neutral copolymer poly(trimethylammonium ethylacrylate methylsulfate)-*b*-poly(acrylamide), used in solutions with a negatively charged surfactant sodium dodecyl sulfate (SDS). The two polymers aforementioned are abbreviated in the following as PANa-*b*-PAM and PTEA-*b*-PAM. The cationic monomer trimethylammonium ethylacrylate is sometimes referred to as [2-(acryloyloxy)ethyl]trimethylammonium in the literature, and it is used with chloride counterions[48,49]. The synthesis of these copolymers is based on the Madix technology which uses the xanthate as chain-transfer agent in the controlled radical polymerization[50,51].

Static and dynamic light scattering experiments were performed on the copolymers aqueous solutions (*i.e.* without surfactant) in order to determine the weight-average molecular weight $M_W$ and the mean hydrodynamic radius $R_H$ of the chains[37]. The molecular weights targeted by the synthesis were 5 000-*b*-30 000 g·mol$^{-1}$ for the anionic-neutral diblock and 11 000-*b*-30 000 g·mol$^{-1}$ for the cationic-neutral one. These values correspond to 69 monomers of acrylic acid, 41 of trimethylammonium ethylacrylate methylsulfate and 420 of acrylamide. The molecular weight of the whole chain as determined from light scattering were found at $M_W^P$ = 43 500 ± 1000 g·mol$^{-1}$ for PANa-*b*-PAM, and $M_W^P$ = 44 400 ± 2000 g·mol$^{-1}$ for PTEA-*b*-PAM[37]. The agreement between the nominal and experimental values is reasonable. The average hydrodynamic radius obtained from dynamic light scattering (value of the quadratic term in the cumulant analysis) is $R_H$ = 55 Å for both copolymers and the polydispersity index 1.6.

Poly(acrylic acid) is a weak polyelectrolyte and its ionization state depends on the pH. In order to derive the molecular weight of the acrylic acid block, titration experiments were performed. By slow addition of sodium hydroxide 0.1 N, the pH of PANa-*b*-PAM solutions was varied systematically from acidic to basic conditions. Titration curves and equivalences on copolymers were compared to those obtained on a PANa homopolyelectrolyte with known molecular weight (2 000 g·mol$^{-1}$). This procedure allows us to get for the anionic block a degree of polymerization $n_{PE}$ = 90 instead of the expected 69, yielding $M_W$(PANa block) = 6 500 g·mol$^{-1}$. In the sequel of the paper, the molecular weight for PANa-*b*-PAM is assumed to be 6 500-*b*-37 000 g·mol$^{-1}$. Poly(trimethylammonium ethylacrylate methylsulfate) is a strong polyelectrolyte which ionization state is pH independent. In such cases, titration experiments are not appropriate. For the interpretation of the neutron data related to the PTEA-*b*-PAM complexes, we will consider the nominal molecular weight 11 000-*b*-30 000 g·mol$^{-1}$ to be the actual one, since within the experimental errors it is close to the one obtained experimentally (44 400 ± 2000 g·mol$^{-1}$).

DTAB and SDS were purchased from Sigma and used without further purification. Both surfactants are with C12 aliphatic chains and exhibit an hexagonal mesophase at high concentrations[52,53]. The critical micellar concentrations in D$_2$O are 0.42 wt. % (15.3 mmol l$^{-1}$) for DTAB[54] and 0.21 wt. % (8.3 mmol l$^{-1}$) for SDS[55].

Mixed solutions of surfactant and polymer were prepared by mixing a surfactant solution to a polymer solution, both prepared at the same concentration c (wt. %) and same pH (pH 7). The relative amount of each component is monitored by the charge ratio Z, $Z = [S]/n_{PE}[P]$ where [S] and [P] are the molar surfactant and polymer concentrations and $n_{PE}$ is the degree of polymerization of the polyelectrolyte block. $Z = 1$ describes the isoelectric solution characterized by the same number densities of positive and negative chargeable ions. In the mixed solutions, the surfactant and polymers concentrations $c_S$ and $c_P$ read :

$$c_S = c Z (z_0 + Z)^{-1} \text{ and } c_P = c z_0 (z_0 + Z)^{-1} \tag{11}$$

with $z_0 = M_W^P / n_{PE} M_W^S$. According to the $M_W^P$'s found for the polymers, $z_0$ = 1.567 for PANa-*b*-PAM/DTAB and $z_0$ = 3.468 for PTEA-*b*-PAM/SDS. The description of the mixed solutions in terms of c and Z is important since it allows the comparison between polymers with different structures and molecular weights. In this work, we focus on dilute solutions and high charge ratios (Z > 1), *i.e.* on solutions where the main component (apart from water) is the surfactant.

## III.2 – Small-Angle Neutron Scattering and Scattering length densities

Small-angle neutron scattering was performed at three different facilities (Argonne National Laboratory, USA; Laboratoire Léon Brillouin and Institute Laue-Langevin, France), and all runs were consistent with each others. Here, only the data obtained on the beam lines D11 and D22 at the Institute Laue-Langevin are shown. Copolymers/surfactant solutions were prepared at a concentration **c** = 1 % using $D_2O$ as a solvent for contrast reasons. On D22, the data collected at 2 m and 14 m cover a range in wave-vector : $1.5 \times 10^{-3}$ Å$^{-1}$ and 0.25 Å$^{-1}$, with an incident wavelength of 12 Å. On D11, three settings were used (1.1, 4.5 and 20 m) with a neutron wavelength of 8 Å and a wave-vector resolution $\Delta q/q$ of 10 %. The spectra are treated according to the standard Institute Laue-Langevin procedures, and the scattering cross sections are expressed in cm$^{-1}$.

The list of the molecular weight and volumes, coherent scattering lengths and length densities of the chemical species studied in this work are given in Tables I - III. We have re-examined these quantities in the light of a more extensive reference survey. The data provided here are slightly different from the one already published by us. For the surfactant micelles, we have computed the scattering length densities assuming that the micellar aggregates are homogeneous spheres (this is a reasonable assumption for the q-range covered here) and that they are made of $N_{S/M}$ elementary scatterers. In the present model, the elementary scatterer comprises one aliphatic chain and β counterions (Br$^-$ for DTAB and Na$^+$ for SDS). β is the ratio (0 < β < 1) of condensed counterions per surfactant molecule.

For poly(acrylic acid) in its charged (as a sodium salt) and in its acidic forms, the apparent molal volumes have been measured consistently by different groups at 47.8 and 33 cm$^3$·mol$^{-1}$ respectively[56,57]. These values were used in Table III for calculating the scattering densities[58-60]. The case of poly(acrylamide) is interesting. It has been shown that for this hydrosoluble polymer, there is an isotope exchange in the $NH_2$ terminal groups when the solvent is deuterated water, or a mixture of deuterated and hydrogenated water. In $D_2O$, this exchange yields at equilibrium a monomer with chemical formula $CH_2CH-COND_2$, instead of $CH_2CH-CONH_2$[60-62]. This exchange was not taken into account in our previous reports. With a molar volume of 53.3 $cm^3 \cdot mol^{-1}$, the scattering length density of the acrylamide monomer[61,62] is $1.86 \cdot 10^{10}$ $cm^{-2}$ in $H_2O$ and $4.19 \cdot 10^{10}$ $cm^{-2}$ in $D_2O$ (Table III). From the scattering length densities listed in Tables I - III, it appears that all the chemical species are contributing to the overall neutron scattering. However, the contributions are of different magnitudes for the polymers and for the surfactants. Poly(acrylamide) in $D_2O$ has for instance a contrast $\Delta\rho = \rho - \rho_S$ which is only ~ 1/3 that of the surfactants. Moreover, in the c and Z ranges put under scrutiny here the total volume of the different monomers (charged and neutral) is also much lower than that of the micelles. As a result, the polymer contribution will be neglected in the fitting of the neutron spectra.

Fig. 3 shows the scattering cross-sections obtained on DTAB (a) and SDS (b) surfactant solutions at c = 1 wt. %. The data are plotted in the Porod representation ($q^4 \times I(q)$ *versus* q) in order to emphasize the signatures of the interfaces between elementary scatterers and solvent molecules. The two Porod intensities exhibit oscillations with a first maximum at 0.15 $Å^{-1}$. The continuous lines through the data points result from best fit calculations (see section IV.1). Figs. 4 and 5 show the neutron scattering cross-sections obtained for PANa-*b*-PAM/DTAB and PTEA-*b*-PAM/SDS solutions in $D_2O$ at **c** = 1 % for Z = 1, 2, 5 and 10. Each spectrum has been shifted with respect to each other for sake of clarity. For these values of the charge ratio, the formation of the mixed aggregates occurs spontaneously through the mechanism of electrostatic self-assembly. All eight neutron intensities are characterized by two features : a strong forward scattering (q → 0), and a structure peak at the wave-vector $q_0$ ~ 0.16 $Å^{-1}$. As highlighted in earlier publications[34,35], these features are the signatures of the core-shell microstructure of the mixed complexes. Our goal here is to provide a quantitative analysis of these spectra, and to retrieve the aggregation numbers and the numbers of polymers per micelles in the aggregates. The description and discussion of the data for Z < 1 can be found in Ref.[35] for PANa-*b*-PAM/DTAB and in Ref.[36] for PTEA-*b*-PAM/SDS.

# IV – Analysis and Discussion

## IV.1 – Surfactant Micelles

In order to interpret the SANS data quantitatively, we model the dodecyltrimethylammonium bromide (DTAB) or the sodium dodecyl sulfate (SDS) micelles as homogeneous, spherical and slightly polydisperse particles. The micelles are made from $N_{S/M}$ surfactants and from $\beta N_{S/M}$ counterions condensed on the surface. For monovalent ionic surfactants with monovalent counterions, the surface charge is $(1-\beta)N_{S/M}e$, where $e$ is the elementary charge. The elementary scatterer (ES) for these ionic micelles comprises thus one aliphatic chain and of $\beta$ counterions. The molecular volume and the coherent scattering lengths are additive quantities :

$$v_0(ES) = v_0(Surf) + \beta v_0(Count), \qquad (12a)$$

$$b_N(ES) = b_N(Surf) + \beta b_N(Count). \qquad (12b)$$

Table I and II recapitulate the different $v_0$ and $b_N$-values for DTAB and SDS in their neutral and ionized forms. The average scattering length densities $\rho_N$ ($\rho_N = b_N/v_0$) are also given. The coefficient $\beta$ was taken from the literature and found to be 0.75 for DTAB[63] and 0.73 for SDS[55]. The SANS data shown in the Porod representation of Figs. 3 were fitted using Eq. 2 and assuming a Gaussian distribution for the micellar radius. The quantities that are deduced are the average micellar radius $\overline{R}$, the polydispersity $s_R = \sigma_R/\overline{R}$ of the distribution and the average number density of micelles $n_{Mic}$ at c = 1 wt. % (Eq. 4). From these data, the number $N_{S/M}$ of surfactants per micelle can be estimated. It can be calculated from the average volume of a micelle using :

$$N_{S/M}^{(1)} = \frac{4\pi}{3v_0} \int_0^\infty R^3 G(R, \overline{R}, \sigma_R) dR, \qquad (13)$$

where the molecular volume $v_0$ of an elementary scatterer follows Eq. 12a. The aggregation number can be determined too from the number density of micelles. At c = 1 wt. %, the concentration of surfactants which are under the form of micelles is c – cmc, yielding :

$$N_{S/M}^{(2)} = 1.11 \frac{c - cmc}{M_W^S} N_A \frac{1}{n_{Mic}} \qquad (14)$$

The exponents (1) and (2) in Eqs. 13 and 14 refer to the different determinations of the aggregation numbers. As shown in Table III, the aggregation numbers are of the order of 50 for both surfactants, with an uncertainty of 5 % for the first determination and of 10 % for the second. For DTAB, $N_{S/M}^{(1)} = 56 \pm 2$ and $N_{S/M}^{(2)} = 50 \pm 5$, whereas for SDS micelles $N_{S/M}^{(1)} = 50 \pm 2$ and $N_{S/M}^{(2)} = 47 \pm 4$. The two determinations are in excellent agreement with each other, as well as with values from the literature. For DTAB at room temperature (T = 25°C), Bales and Zana have performed recently an extended survey of the physico-chemical properties of this surfactant, and found for DTAB micelles (in $H_2O$) an aggregation number $N_{S/M}(DTAB) = 54.7 \pm 1.6$[63]. For SDS micelles, Bales and coworkers have compared aggregation numbers determined by different techniques such as light scattering, ultracentrifugation, SANS[64] and time resolved fluorescence quenching[55]. The agreement between these various methods proves that the SANS measurements presented here are reliable and that the microscopic parameters used to describe the micelles are satisfactory. In the sequel of the paper, we adopt the following values for the aggregation numbers, $N_{S/M}(DTAB) = 53$ and $N_{S/M}(SDS) = 50$.

## IV.2 – Colloidal Complexes

*IV.2.1 –Number of Micelles and Polymers per Colloidal Complex*

Let us denote $N_{M/C}$ and $N_{P/C}$ the average numbers of micelles and polymers per colloidal complex respectively, and r the ratio between these two numbers ($r = N_{M/C}/N_{P/C}$). In order to model the scattering properties of the mixed aggregates, the following assumptions are made :

    1 – The aggregation numbers $N_{S/M}$ (number of surfactants per aggregate) for the micelles located in the cores are identical to the values found in dilute solutions.

    2 – At large Z (Z > 5), the polymer is the minority component. Since the polymer scattering contrast is weak when compared to that of the surfactant, we suppose that the scattering of complexes arise only from the micelles comprised in the cores.

    3 – At large Z (Z > 5), all the diblocks participate to the formation of the colloidal complexes. This assumption is in agreement with several recent studies[65-70].

Fig. 6 shows the neutron intensities obtained for PANa-*b*-PAM/DTAB and PTEA-*b*-PAM/SDS solutions at Z = 5 and Z = 10, together with the data obtained with the surfactant alone (c = 1 wt. %, Z = ∞). The data are those of Figs. 3 – 5. At high wave-vectors (q > 0.05 Å$^{-1}$), the intensities observed for the mixed solutions exhibit a q-dependence that is reminiscent of that of the surfactants. A multiplicative coefficient, noted α has been applied to the surfactant data in order to demonstrate the superimposition of the scattering functions in this range. Note that for PTEA-*b*-PAM/SDS at Z = 10, the agreement with SDS is remarkable. The data of Fig. 6 suggest that at large Z, the solutions can be seen as a coexistence state comprising surfactant micelles and colloidal complexes. The scattering cross-section then expresses as the sum of two terms, one arising from micelles incorporated into the complexes (Eq. 7) and one arising from the free micelles (Eq. 1) :

$$\frac{d\sigma}{d\Omega}(q,c,Z) = V^2 \Delta\rho^2 \, F(q,R) \left[ n_{Mic}^{(1)}(c,Z) \langle S_C(q,N) \rangle_{MonteCarlo} + n_{Mic}^{(2)}(c,Z) \right] \quad (15)$$

$n_{Mic}^{(1)}(c,Z)$ and $n_{Mic}^{(2)}(c,Z)$ are the number densities of micelles in the associated and free state respectively. These numbers obey moreover the conservation laws :

$$n_{Mic}^{(1)}(c,Z) + n_{Mic}^{(2)}(c,Z) = \frac{c_S(c,Z) - cmc}{M_W^S} \mathcal{N}_A \frac{1}{N_{S/M}} \quad \text{and} \quad (16a)$$

$$n_{Mic}^{(2)}(c,Z) = \alpha \frac{c(=1wt.\%) - cmc}{M_W^S} \mathcal{N}_A \frac{1}{N_{S/M}} \quad (16b)$$

where $c_S$ is given in Eq. 11. Eq. 16b introduces the coefficient α and expresses the superposition of the intensities observed in Fig. 6. For simplicity, Eq. 15 is written for monodisperse particles and aggregates. Experimental values for α, $n_{Mic}^{(1)}$, $n_{Mic}^{(2)}$ and r are shown in Table V for Z = 5 and Z = 10 and for the two polymer-surfactant systems. Note here that these determinations are based on a minimum set of assumptions. More specifically, it is based on the principle of superposition of the scattering curves at different Z and on the knowledge of a unique parameter, the aggregation number for the micelles $N_{S/M}$. We find that r = $N_{M/C}/N_{P/C}$ is of the order 1 for PANa-*b*-PAM/DTAB and of the order 1/3 for PTEA-*b*-PAM/SDS. r = 1/3 means that the core is made in average by three polymers for one SDS micelle. In Table V are also estimated the effective charge ratios within the aggregates noted $Z_{eff}$, as determined from the r's. As anticipated, the effective charge ratios $Z_{eff}$ are very different from the Z at which the solution mixing was performed. $Z_{eff}$ is found to be 0.66 ± 0.06 for PANa-*b*-PAM/DTAB and 0.38 ± 0.02 for PTEA-*b*-PAM/SDS. Effective charge ratios below unity suggest that the complexation occurs with an excess of charges coming from the polyelectrolyte chains, and that overcharging might play a role in the formation of the mixed colloids.

*IV.2.2 – Distributions of Aggregation Numbers*

Quantitative fits to the scattering data are obtained using Eq. 10. This equation presupposes that micelles and aggregates are only slightly polydisperse, and that under such approximation the integral over the two distributions can be treated separately. For Z = 5 and 10, the contribution of the free micelles is taken into account and added to the overall intensity. Note that for the calculation of the average interparticle structure factor $\langle S_C(q,N) \rangle_{MonteCarlo}$, we have considered that the volume fraction of micelles $\phi_C$ inside the core and defined by Eq. 5 is 0.5. The results of the fitting are displayed as continuous lines in Fig. 4 and 5. With this model, we are able to account accurately for the overall scattering intensity, and especially for the structure peak located around 0.16 Å$^{-1}$. The agreement obtained with PANa-*b*-PAM/DTAB is excellent, whereas for PTEA-*b*-PAM/SDS at Z = 1 and 2 it is slightly less good in the high q-range. Explanations for these discrepancies are proposed in the conclusion section. Finally, this approach allows us to derive the distribution of aggregation numbers (that is the distribution of numbers of micelles located in the aggregates) for the two systems. These distributions are shown in Fig. 7a and 7b, respectively. For PANa-*b*-PAM/DTAB, the distribution shifts to higher aggregation numbers (from 100 to 230) and broadens with increasing Z. For PTEA-*b*-PAM/SDS, the distribution remains centered around $\overline{N}_{M/C}$(SDS) = 70, with the standard deviation that remains unchanged. The average aggregation numbers $\overline{N}_{M/C}$ and $\overline{N}_{M/C}$ derived from this approach are listed in Table VI, together with the radius R$_C$ and with the polydispersity s$_{RC}$. To our knowledge, this is the first time that colloidal complexes resulting from electrostatic self-assembly have been described in terms of their separate aggregation numbers (here micelles and polymers) and their distributions.

# V – Concluding Remarks

In the present paper, we have re-examined neutron scattering data that were obtained on colloidal complexes resulting from the self-assembly between charged-neutral copolymers and oppositely charged surfactants[34-37,40]. Two block copolymers/surfactant systems are highlighted, on one hand

poly(acrylic acid)-*b*-poly(acrylamide) with dodecyltrimethylammonium bromide and on the other hand poly(trimethylammonium ethylacrylate methylsulfate)-*b*-poly(acrylamide) with sodium dodecyl sulfate. Our goal was to use the quantitative scattering cross-sections offered by SANS to derive the aggregation numbers and distributions for these novel core-shell structures.

As for the neutron data, we are now able to justify one important hypothesis. Although the aggregates result from the association between copolymers and micelles, it is sufficient to take into account only the micelles for fitting the SANS data. One reason is that there is an isotope exchange in the terminal groups of acrylamide monomers in $D_2O$ which considerably decreases the scattering contrast. The scattering length density of the acrylamide monomer passes from $1.86 \cdot 10^{10}$ cm$^{-2}$ in $H_2O$ to $4.19 \cdot 10^{10}$ cm$^{-2}$ in $D_2O$. Another reason is that at large Z, the surfactant component is in large excess with respect to the polymers. In this effort of being quantitative, we have also revisited our data on the DTAB and SDS surfactant micelles. We now provide sizes and aggregation numbers for the micelles that are consistent with literature data[55,63]. The agreement between the different evaluation techniques proves that the microscopic parameters used to model the micellar aggregates are satisfactory. We have used throughout the paper the values $N_{S/M}(DTAB) = 53$ and $N_{S/M}(SDS) = 50$.

For the polymer-surfactant complexes, we have developed two complementary approaches. The first one is based on the principle of superposition of the scattering curves at different Z. To do so, we need not formulate any hypothesis on the structure and composition of the complexes. Instead, we have used an experimental intensity (that of surfactant) to scale the amount of unassociated micelles that are present at large Z, and to deduce the amount of surfactants incorporated in the complexes. This provided us with the average number of micelles per polymer $r = N_{M/C}/N_{P/C}$ in the mixed aggregates. We found that r is of the order 1 for PANa-*b*-PAM/DTAB and of the order 1/3 for PTEA-*b*-PAM/SDS. r = 1/3 means that the core is made in average by 3 polymers for one SDS micelle. From these values, the effective charge ratios within the aggregates were estimated and found at $Z_{eff} = 0.66 \pm 0.06$ for PANa-*b*-PAM/DTAB and $Z_{eff} = 0.38 \pm 0.02$ for PTEA-*b*-PAM/SDS.

Values for the effective charge ratios below unity mean that the complex formation is not accompanied by a full compensation of the positive and negative charges coming from the two components. In the two instances considered here, the amount of polyelectrolytes needed to build the core-shell structures always exceeds the number necessary to balance the charge of the micelles. This is true for the polyelectrolyte block that has less charges than the micelle ($n_{PE} < N_{S/M}$ for PTEA-*b*-PAM/SDS), as well as for the block that has more charges than the micelle ($n_{PE} > N_{S/M}$ for PANa-*b*-PAM/DTAB). This result suggests that the electrostatic self-assembly in the present polymer-surfactant systems is accompanied by the overcharging of the micelles. In this context, we recall that the predictions based on a pure counterion release model would give a charge ratio for the mixed aggregates $Z_{eff} = 1$, as it was observed for the DNA-liposome system[8].

Finally, the SANS intensities at Z = 1, 2, 5 and 10 were fitted successfully for the two systems using Eq. 10, yielding the distribution functions for the core radius as well as for the aggregation numbers (Figs. 7). We had to fix the volume fraction $\phi_C$ for micelles in the cores at 0.5 in order to reproduce the structure peak at $q_0 \sim 0.16$ Å$^{-1}$ in its position and amplitude. Such a high volume fraction justifies the description of the core in terms of a dense coacervate microphase. In PANa/DTAB precipitates where PANa is an homopolyelectrolyte, the volume fraction calculated from the *Pm3n* cubic structure is 0.524[37]. For the PTEA-*b*-PAM/SDS system at Z = 1 and 2, there is a slight disagreement between the calculated and the experimental intensities. One explanation could be the presence of methylsulfate counterions in the core (compensating then the excess of positive charges in the core). Because of their three hydrogen atoms, these methylsulfate might change the neutron contrast inside the core and the assumption that only the SDS micelles contribute to the scattering would not be valid. Another possibility for this discrepancy would be that at the high weight concentrations reached in the cores, the SDS micelles become slightly anisotropic. Having larger aggregation numbers would shift the position of the structure peak to the left on the wave-vector scale, as shown in Fig. 5.

**Acknowledgements** :

I would like to thank Mathias Destarac, Ronny Eng, Isabelle Grillo, Mikel Morvan, Julian Oberdisse, Ralph Schweins, Amit Sehgal for their help and support during the course of this research. I am specially grateful to Julian Oberdisse for the Monte Carlo simulations of the core structures, and to François Schosseler for the references on the scattering lengths densities of acrylamide and acrylic acid monomers. The Laboratoire Léon Brillouin (Saclay, France) and the Institute Laue-Langevin are acknowledged for their technical and financial support. This research is supported by Rhodia Inc. and by the Centre de la Recherche Scientifique in France.

# Tables and Table Captions

| species | $v_0$ Å$^3$ | $b_N$ $10^{-12}$ cm | $\rho_N$ $10^{10}$ cm$^{-2}$ |
|---|---|---|---|
| DTAB | 495.5 | -1.138 | - 0.23 |
| DTA$^+$ | 456.2 | -1.817 | - 0.40 |
| Br$^-$ | 39.3 | 0.679 | + 1.73 |
| [DTA$^+$]+0.75×[Br$^-$] | 485.7 | -1.308 | - 0.27 |

**Table I** : Specific molecular volume ($v_0$), scattering length ($b_N$) and scattering length density ($\rho_N$) for dodecyltrimethylammonium bromide (DTAB, molecular weight $M_W^S$ = 308.35 g·mol$^{-1}$). The last line corresponds to the values assuming that DTAB micelles in water are made of $N_{S/M}$ elementary scatterers, each scatterer consisting of one DTA$^+$ and 3/4 Br$^-$. This definition aims to take into account the condensation of the counterions on the micelle surface[63].

| species | $v_0$ (Å$^3$) | $b_N$ ($10^{-12}$ cm) | $\rho_N$ ($10^{10}$ cm$^{-2}$) |
|---|---|---|---|
| SDS | 412.2 | +1.595 | + 0.39 |
| SD$^-$ | 403.1 | +1.232 | + 0.31 |
| Na$^+$ | 9.1 | 0.363 | + 3.99 |
| [SD$^-$]+0.73×[Na$^+$] | 409.7 | 1.497 | + 0.37 |

**Table II** : Same quantities as in Table I for sodium dodecyl sulfate (SDS, molecular weight $M_W^S$ = 288.38 g·mol$^{-1}$). The last line corresponds to values assuming that 73 % of the sodium counterions on the micelle surface[55].

| species | chemical formula | $M_W$ g·mol$^{-1}$ | $V_{mol}$ cm$^3$·mol$^{-1}$ | $v_0$ Å$^3$ | $b_N$ 10$^{-12}$ cm | $\rho_N$ 10$^{10}$ cm$^{-2}$ |
|---|---|---|---|---|---|---|
| acrylic acid | CH$_2$CH-COOH | 72.06 | 47.8 | 79.4 | 1.66 | + 2.09 |
| sodium acrylate | CH$_2$CH-COO$^-$,Na$^+$ | 94.04 | 33 | 54.8 | 2.40 | + 4.37 |
| trimethylammonium ethylacrylate | CH$_2$CH-COO(C$_2$H$_4$)-N$^+$(CH$_3$)$_3$ | 158.22 | nd | nd | 1.42 | nd |
| acrylamide in H$_2$O | CH$_2$CH-CONH$_2$ | 71.08 | 53.3 | 88.6 | 1.64 | + 1.856 |
| acrylamide in D$_2$O | CH$_2$CH-CONHD$_2$ | 73.08 | 53.3 | 88.6 | 3.71 | + 4.191 |
| deuterated water | D$_2$O | 20.02 | 18.0 | 30 | 1.92 | + 6.38 |

**Table III :** Chemical formula, molecular weight ($M_W$), molar volume ($V_{mol}$), molecular volume ($v_0$), coherent neutron scattering length ($b_N$) and length density ($\rho_N$) of the polymers studied in this work.

| Surfactant | cmc mmol·l$^{-1}$ | $\bar{R}_{Mic}$ Å | $s_R$ | $n_{Mic}$ cm$^{-3}$ | $N_{S/M}^{(1)}$ | $N_{S/M}^{(2)}$ |
|---|---|---|---|---|---|---|
| DTAB | 15.3$^a$ | 18.15 | 0.169 | 2.5·10$^{17}$ | 56 ± 2 | 50 ± 5 |
| SDS | 8.3$^b$ | 16.90 | 0.170 | 3.9·10$^{17}$ | 54 ± 2 | 47 ± 4 |

**Table IV** : Parameters derived from the fitting of the DTAB-D$_2$O and SDS-D$_2$O scattering intensities measured at c = 1 wt. %. $N_{S/M}^{(1)}$ and $N_{S/M}^{(2)}$ are obtained from Eqs. 13 and 14 with an uncertainty of ∼ 5 % and ∼ 10 %, respectively. These values are in excellent agreement with literature data (see text). Throughout the paper we adopt the following values for the aggregation numbers, $N_{S/M}$(DTAB) = 53 and $N_{S/M}$(SDS) = 50. $^a$ : Ref.[52,54]; $^b$ : Ref.[53,55].

| system | Z | α | $n_{Mic}^{(1)}$ $10^{16}$cm$^{-3}$ | $n_{Mic}^{(2)}$ $10^{16}$cm$^{-3}$ | $r = N_{M/C}/N_{P/C}$ | $Z_{eff}$ |
|---|---|---|---|---|---|---|
| PANa-*b*-PAM/DTAB | 5 | 0.42 | 3.7 | 9.9 | 1.0 | 0.59 |
|  | 10 | 0.64 | 2.7 | 15.1 | 1.3 | 0.74 |
| PTEA-*b*-PAM/SDS | 5 | 0.43 | 2.1 | 16.0 | 0.33 | 0.40 |
|  | 10 | 0.65 | 1.2 | 23.8 | 0.29 | 0.35 |

**Table V** : Parameters deduced from the comparison of the intensities obtained for polymer-surfactant complexes (c = 1 wt. %, Z = 5 and 10) and for surfactants alone (c = 1 wt. %). $n_{Mic}^{(1)}$ and $n_{Mic}^{(2)}$ denote the number densities of micelles located in the complexes and free in solution (*i.e.* unassociated), respectively. r is the number of micelles per polymer in the complexes and $Z_{eff}$ is the effective charge ratio derived from r ($Z_{eff} = N_{S/M} \cdot r/n_{PE}$).

| system | Z | $R_C$ (Å) | $s_{RC}$ | $\overline{N}_{M/C}$ | $\overline{N}_{P/C}$ |
|---|---|---|---|---|---|
| PANa-*b*-PAM/DTAB | 1 | 108 | 0.16 | 106 | ~ 100 |
|  | 2 | 124 | 0.14 | 157 | ~ 160 |
|  | 5 | 140 | 0.18 | 232 | 232 |
|  | 10 | 140 | 0.18 | 232 | 184 |
| PTEA-*b*-PAM/SDS | 1 | 88 | 0.17 | 71 | ~ 240 |
|  | 2 | 80 | 0.24 | 55 | ~ 180 |
|  | 5 | 86 | 0.2 | 69 | 212 |
|  | 10 | 92 | 0.16 | 79 | 274 |

**Table VI** : List of parameters obtained from the fitting of the neutron scattering cross-sections using Eq. 10. $R_C$ and $s_{RC}$ characterize the Gaussian distribution in radius, $R_C$ being the center of the distribution and $s_{RC}$ the polydispersity. $\overline{N}_{M/C}$ and $\overline{N}_{P/C}$ are the average numbers of micelles and polymers found in complexes. At Z = 1 and Z = 2, $\overline{N}_{P/C}$ is estimated using r = 1 for the PANa-*b*-PAM/DTAB and 0.3 for PTEA-*b*-PAM/SDS.

# Figure Captions

**Figure 1 :** Representation of a colloidal complex resulting from the self-assembly of oppositely charged block copolymers and surfactants. Depending on molecular weight, the radius of the core ranges from 10 to 20 nm and the corona thickness between 10 and 50 nm.

**Figure 2 :** Illustration of the two dispersion states discussed in the theory section. Left : Dispersed small spheres with radius R. Right : dispersed aggregates made from the association of N small spheres. The scattering cross-sections corresponding to each state are given by Eq. 1 and 7, respectively.

**Figure 3 :** Porod representation ($q^4 \times d\sigma(q)/d\Omega$ *versus* q) of the scattering cross-sections obtained on DTAB (a) and SDS (b) surfactant solutions at c = 1 wt. %. The continuous lines result from best fit calculations as explained in the text.

**Figure 4 :** Neutron scattering cross-sections obtained from aqueous solutions containing charged-neutral copolymers and surfactants. The total concentration is c = 1 wt. % and the temperature is 25 °C. The polymers used are the poly(sodium acrylate)-*b*-poly(acrylamide) diblocks and the surfactant is dodecyltrimethylammonium bromide (DTAB). Poly(sodium acrylate) is negatively charged in water, poly(acrylamide) is neutral and DTAB is positively charged. The molecular weights for the charged and for the neutral blocks are 6500 and 37000 g mol$^{-1}$, respectively. The different scattering curves were measured at different charge ratios Z between the surfactants and the polymers (Z = 1, 2, 5 and 10). The continuous curves are fits assuming for the assembly the microstructure outlined in Figs. 1 and 2. Each spectrum has been shifted with respect to each other for sake of clarity.

**Figure 5 :** Same as in Figure 4, but for the cationic-neutral copolymer poly(trimethylammonium ethylacrylate methylsulfate)-*b*-poly(acrylamide) mixed in solutions with a negatively charged sodium dodecyl sulfate (SDS). The molecular weights for the charged and for the neutral blocks are 11 000 and 30000 g mol$^{-1}$, respectively.

**Figure 6 :** Neutron scattering intensities obtained for PANa-*b*-PAM/DTAB and PTEA-*b*-PAM/SDS solutions at large Z (c = 1 wt. %), together with data obtained from the pure surfactant system (c = 1 wt. %, Z = ∞) : a) PANa-*b*-PAM/DTAB and Z = 5; b) PANa-*b*-PAM/DTAB and Z = 10; c) PTEA-*b*-PAM/SDS, Z = 5; d) PTEA-*b*-PAM/SDS, Z = 10. A multiplicative coefficient α (in parenthesis) has been applied to the surfactant cross-sections in order to shift the intensity at the level of the mixed solutions. This is to show the coexistence between surfactant micelles and colloidal complexes at large Z.

**Figure 7 :** Distribution functions of aggregation numbers used to fit the neutron scattering data in Fig. 4 and 5 : a) PANa-*b*-PAM/DTAB and b) PTEA-*b*-PAM/SDS. The averages values and polydispersity are listed in Table VI for the micelles and for the polymers.

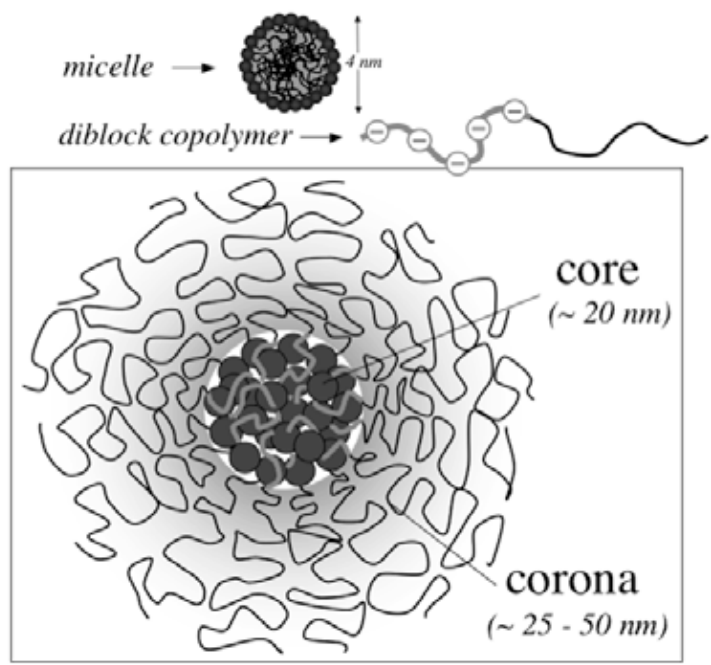

Figure 1

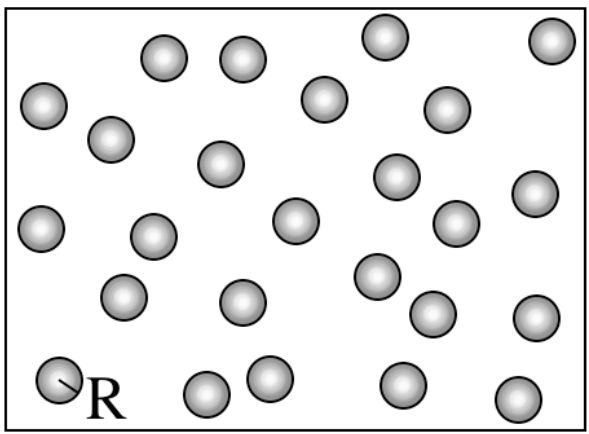 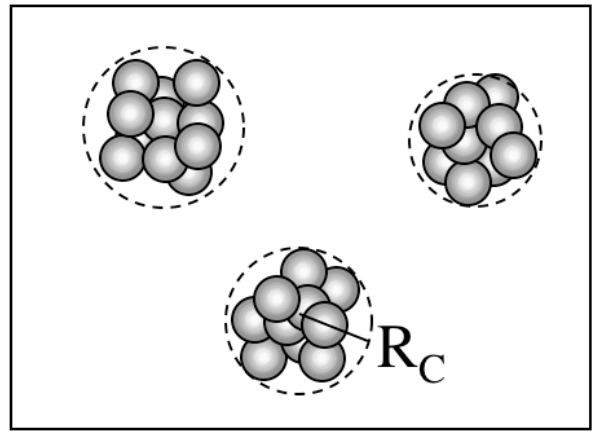

dispersed spheres                dispersed clusters of spheres

Figure 2

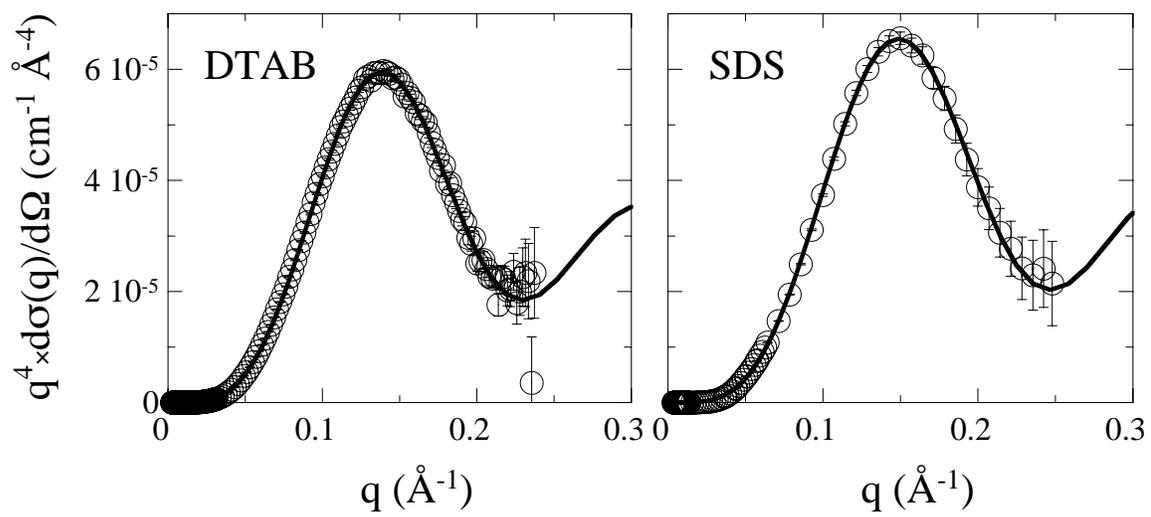

Figure 3

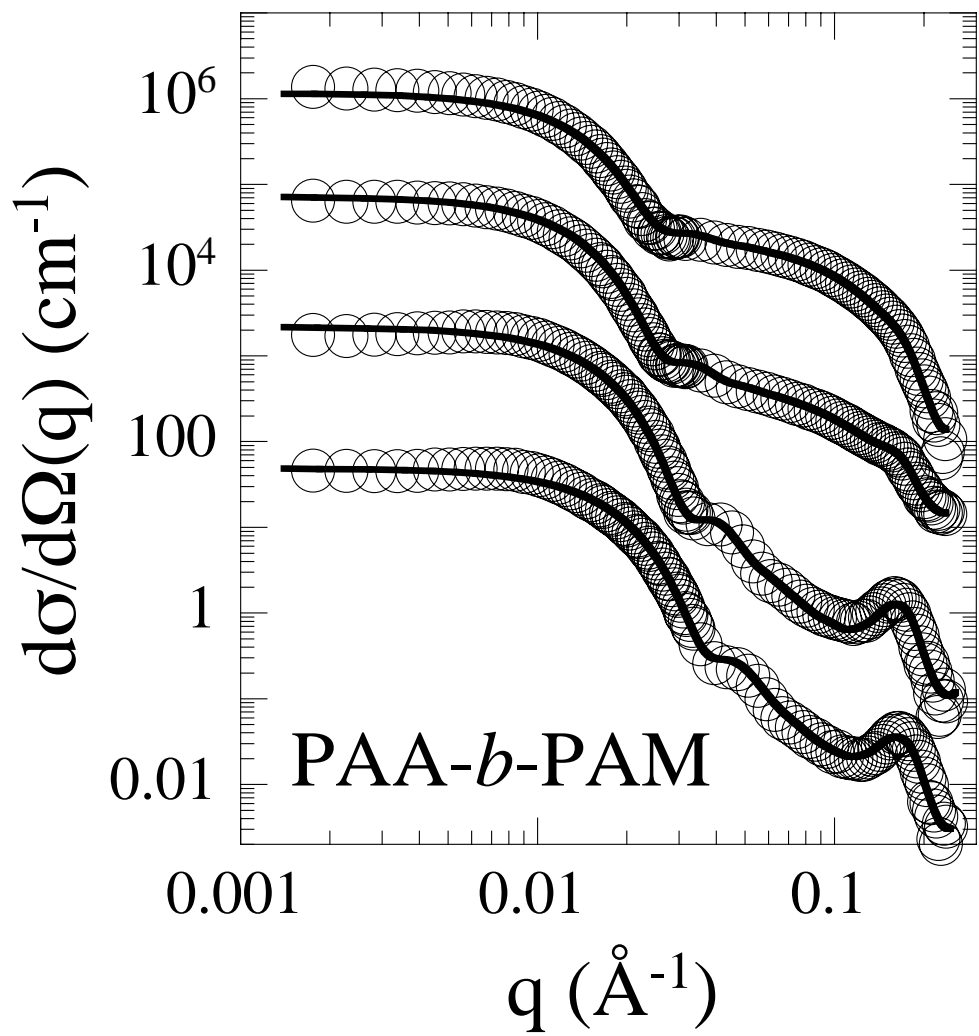

Figure 4

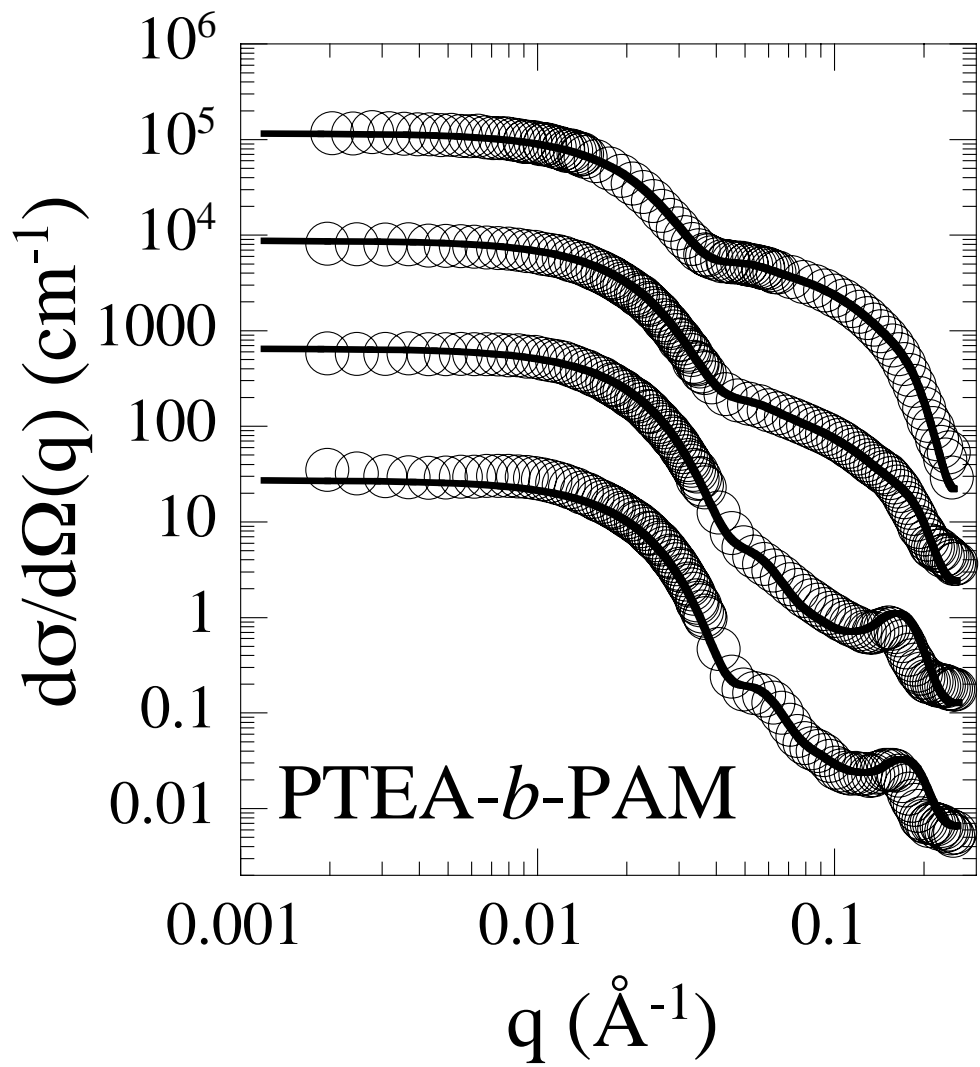

Figure 5

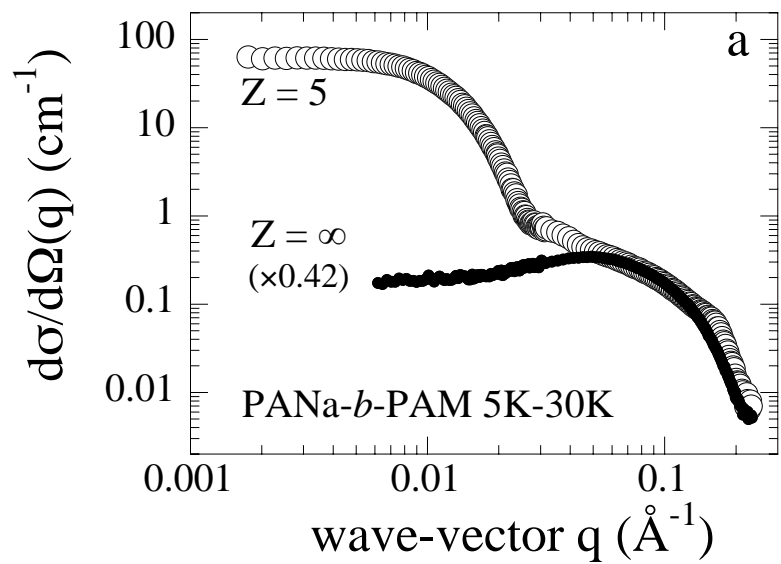

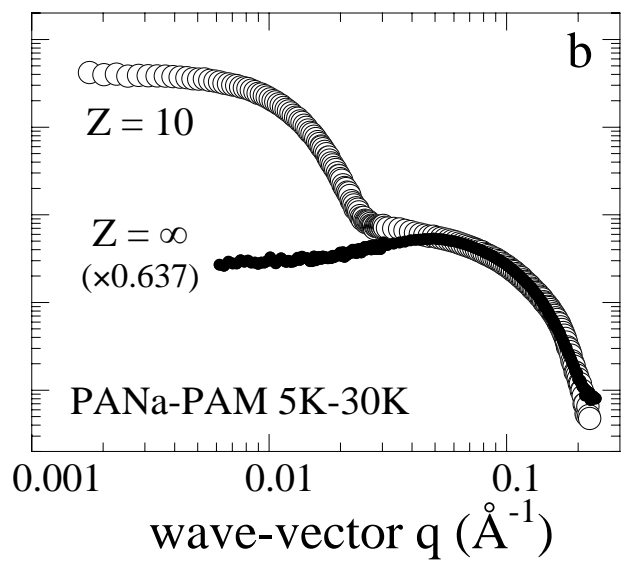

Figure 6a and 6b

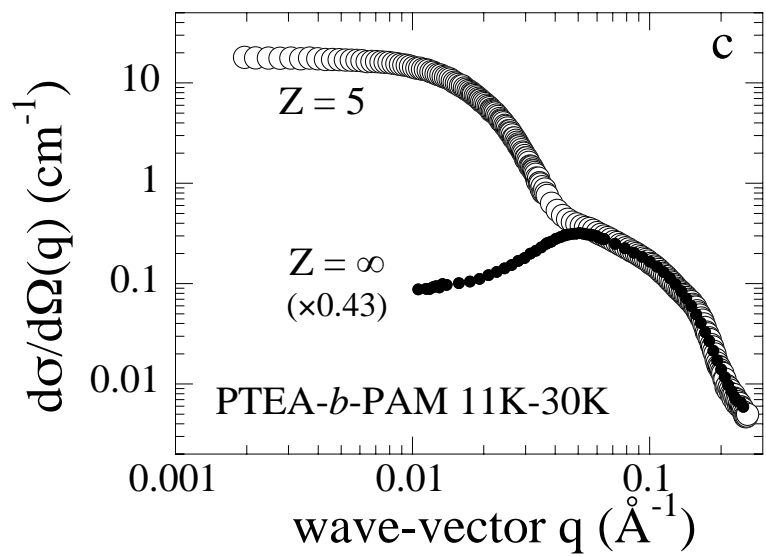

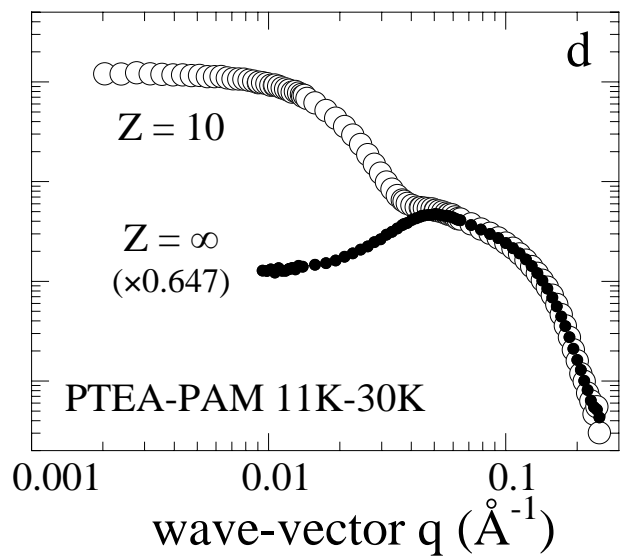

Figure 6c and 6d

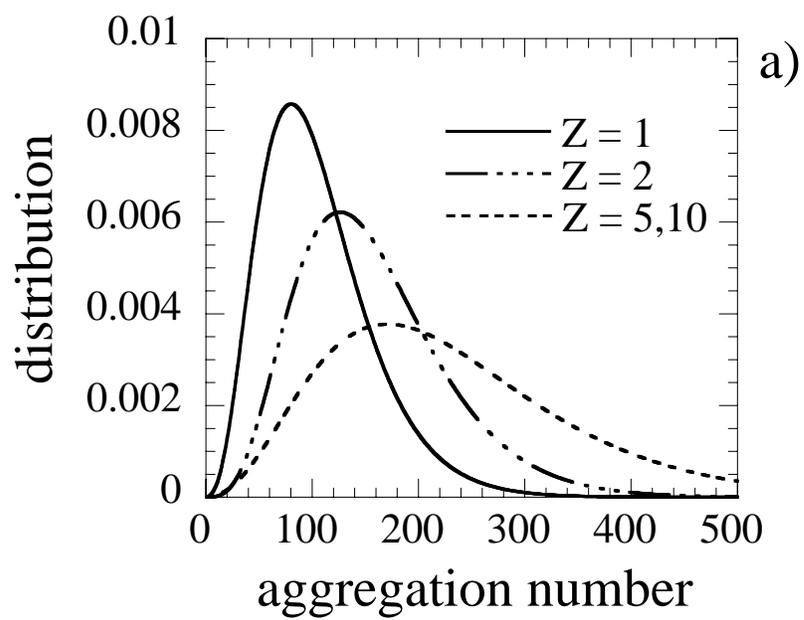
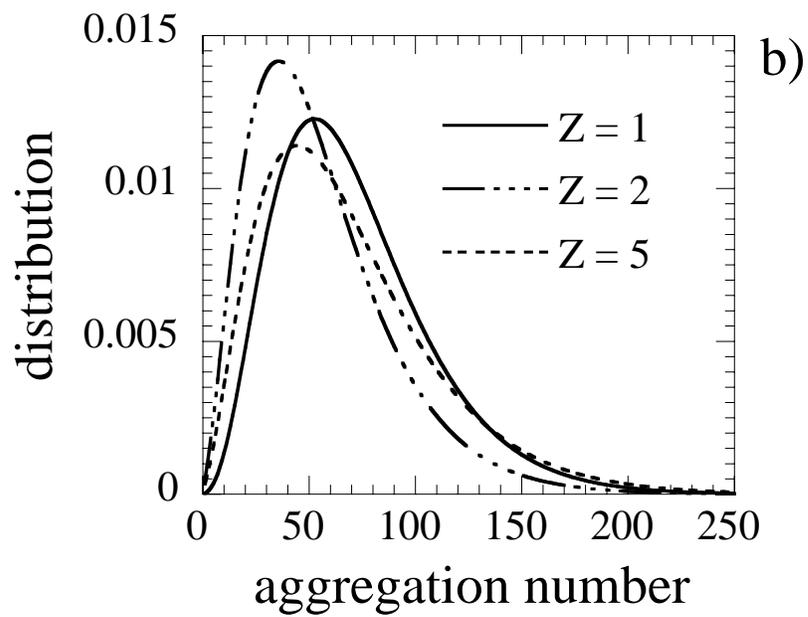

Figure 7